\documentclass[12pt]{article} 
\usepackage[dvips]{graphicx}
\title{The Variance of Standard Option Returns}
\author{Adi Ben-Meir, Jeremy Schiff 
\\ Department of Mathematics,
\\ Bar-Ilan University,
\\ Ramat Gan, 52900, Israel
\\ schiff@math.biu.ac.il }

\begin{document}
\maketitle
\begin{abstract}
The vast majority of works on option pricing operate on the assumption 
of risk neutral valuation, and consequently focus on the expected 
value of option returns, and do not consider risk parameters, such as variance. 
We show that it is possible to give explicit formulae for the variance
of European option returns (vanilla calls and puts, as well as barrier
options), and that for American options the variance can be computed
using a PDE approach, involving a modified Black-Scholes PDE. We show 
how the need to consider risk parameters, such as the variance, and 
also the probability of expiring worthless (PEW), arises naturally for 
individual investors in options. Furthermore, we show that a volatility 
smile arises in a simple model of risk-seeking option pricing.  
\end{abstract}

\section{Introduction} 

Options are utilized by different investors for different goals. On 
the one hand options are often used for hedging, eliminating the risk
of extreme downside events at the expense of forfeiting (some of) the gain 
of upside events. On the other hand, options are used for 
speculation, to increase the leverage of an investor that has strong
convictions about the direction of the market and is willing to expose
herself to substantial risk, with the potential of very large gains 
if the scenarios she anticipates are realized. Other uses are discussed in 
\cite{merton1978returns}-\cite{merton1982returns}. 

Regardless of the strategy in which they are being used,
an option should be considered as a risky asset, with a distribution of 
returns determined by the distribution of returns of the 
underlying asset, but  typically far more extreme, usually with
significant probability of total loss or a gain of hundreds of percent. 
In this context it is bizarre that the most common piece of quantitative data 
given about options is the Black-Scholes price, which represents the 
current value of the {\em expected value} of the return of the option (under 
suitable assumptions about the distribution of returns of the underlying asset). 
Given that the distribution of the returns of the option is extreme, 
knowing its expected value is of limited significance, except maybe for the 
largest investors who are making many purchases and sales, and even these
investors should be interested in knowing the relevant law of large
numbers for their situation, and must take into account strong correlations. 
More information about the distribution of option returns is surely needed in 
order to make rational investment decisions. 

Rather remarkably, we have not succeded in finding any 
literature examining the full distribution of the returns from options, 
even on the assumption of a lognormal distribution of returns of the underlying asset, 
beyond acknowledgment that this distribution is extreme, and, of course,
calculation of its expected value. In part this can be attributed to the fact 
that seminal papers such as  Black and Scholes 
\cite{black1973pricing}, Merton \cite{merton1973theory}, 
Cox, Ross and Rubinstein \cite{cox1979option}, all emphasize  
that any deviation of an option price from its
risk-neutral valuation creates an arbitrage opportunity that cannot exist in 
a rational market. In addition, much literature focuses on the variation  of 
returns from options due to lack of complete knowledge about the distribution of 
underlying assets, their rates of return and (particularly) their volatilities, 
see for example Coval and Shumway \cite{coval2001expected}, Liu and Pan 
\cite{liu2003dynamic} and Goyal and Saretto \cite{goyal2009cross}. 
It is not the intention of this paper to question the rationality
assumptions behind risk-neutral option valuation. The thrust of this paper
is to show that mathematically it is possible to compute different parameters 
describing the distribution of returns from an option, assuming 
a lognormal distribution of returns of the underlying asset, with known parameters. 
We focus on two (related) measures of risk: the variance and the probability of expiring worthless (PEW). 
In the case of vanilla European call and put options these parameters are straightforward
to compute analytically, and we do this in section 2. Analytic expressions can 
also be given for European options with barriers and we illustrate this in section 3. 
For American options, things are more complicated. Analytic expressions cannot be 
found, but the moments of the distribution from which the variance and PEW can 
be derived satisfy {\em modified} Black-Scholes partial differential equations (PDEs) 
giving numerical schemes for computation. This is described in section 4. 

In section 5 we give some simple examples of the use of the risk parameters we 
compute in individual investment decisions. In section 6, we use the variance 
parameter to build a simple model describing risk-averse or risk-seeking market pricing 
of options; we show that such a model is equivalent to the standard model with
a volatility smile. Finally, section 7 contains closing comments and suggestions 
for further research. 

\section{European options}   

Assuming an underlying with a lognormal distribution with parameters 
$r,\sigma$ and current price $S_0$, the current value of the return on 
European call and put options with strike $K$ at time $T$ are
\begin{eqnarray*}
C  &=&  e^{-rT}\left(S_0e^{\left(r-\frac12\sigma^2\right)T+\sigma\sqrt{T}Z} - K\right)_+  \ ,\\
P  &=&  e^{-rT}\left(K-S_0e^{\left(r-\frac12\sigma^2\right)T+\sigma\sqrt{T}Z}\right)_+ 
\end{eqnarray*} 
where in both formulas $Z$ is a standard normal variable. 
A simple computation gives the expected values 
\begin{eqnarray*}
{\bf E}[C]  &=&  S_0 \Phi(-d+\sigma\sqrt{T}) - Ke^{-rT} \Phi(-d)  \ ,\\ 
{\bf E}[P]  &=&  Ke^{-rT} \Phi(d)  - S_0 \Phi(d-\sigma\sqrt{T})
\end{eqnarray*} 
where
$$ d = \frac1{\sigma\sqrt{T}}\ln\left(\frac{Ke^{-rT}}{S_0} \right)  + \frac12 \sigma\sqrt{T}   \ . $$
These formulae are the Black-Scholes prices for calls and puts respectively. Note
both $u={\bf E}[C]$ and $u={\bf E}[P]$ satisfy the Black-Scholes PDE 
\begin{equation}
\frac{\partial u}{\partial T} = 
\frac12\sigma^2 S_0^2 \frac{\partial^2 u}{\partial S_0^2}
+ r S_0 \frac{\partial u}{\partial S_0} - r u 
\ .
\label{BS1}\end{equation}

It is a straightforward calculation to find the variance of $C$ and $P$. We find 
\begin{eqnarray*}
{\bf E}[C^2] &=&  
S_0^2 e^{\sigma^2 T} \Phi(-d+2\sigma\sqrt{T}) 
- 2 K S_0 e^{-rT} \Phi(-d+\sigma\sqrt{T}) 
+ K^2 e^{-2rT} \Phi(-d) \ ,
\\ 
{\bf E}[P^2] &=&  K^2 e^{-2rT} \Phi(d) - 2 K S_0 e^{-rT} \Phi(d-\sigma\sqrt{T}) 
+ S_0^2 e^{\sigma^2 T} \Phi(d-2\sigma\sqrt{T}) \ . 
\end{eqnarray*}
The variances are then computed using ${\rm Var}(C) = {\bf E}[C^2]-\left({\bf E}[C]\right)^2$ and
${\rm Var}(P) = {\bf E}[P^2]-\left({\bf E}[P]\right)^2$.  By the Feynmann-Kac formula, 
both $v={\bf E}[C^2]$ and $v={\bf E}[P^2]$ satisfy the {\em modified} Black-Scholes PDE 
\begin{equation}
\frac{\partial v}{\partial T} = 
\frac12\sigma^2 S_0^2 \frac{\partial^2 v}{\partial S_0^2}
+ r S_0 \frac{\partial v}{\partial S_0} - 2 r v 
\ .
\label{BS2}\end{equation}
The factor $2$ before the last term arises as the discount factor in $C^2$ and $P^2$ is 
$e^{-2rT}$. The relevant final condition and boundary conditions for the modified Black-Scholes PDE
(for both call and put) 
are obtained be requiring that the variance should vanish as the final time is approached,
and as $S_0\rightarrow 0, \infty$. 

In greater generality it is also possible to compute formulae for the $n$th moments ${\bf E}[C^n]$ and 
${\bf E}[P^n]$, $n=3,4,\ldots$. The $n$th moments satisfy the modified Black-Scholes PDE
$$
\frac{\partial v}{\partial T} = 
\frac12\sigma^2 S_0^2 \frac{\partial^2 v}{\partial S_0^2}
+ r S_0 \frac{\partial v}{\partial S_0} - n r v 
$$
due to the discount factor $e^{-nrT}$. There does not appear to be an explicit closed form for the 
moment generating functions of $C$ and $P$, but it can be shown that these also satisfy a suitably 
modified Black-Scholes PDE. 

Both the random variables $C$ and $P$ have a mixed discrete-continuous distribution, with a finite
probability of being zero, and a continuous range of values above zero. For such distributions there
are two possible ways of defining the zeroth order moment, either as $1$ or as the probability to be
nonzero. For options, the probability of expiring worthless (PEW) is a quantity of interest. 
Clearly we have 
\begin{eqnarray*}
{\rm PEW}(C) &=&  \Phi(d) \ , \\ 
{\rm PEW}(P) &=&  \Phi(-d)  \ .
\end{eqnarray*}
Thanks to the relation of these quantities with zeroth order moments, 
they too solve a modified Black-Scholes PDE 
\begin{equation}
\frac{\partial w}{\partial T} = 
\frac12\sigma^2 S_0^2 \frac{\partial^2 w}{\partial S_0^2}
+ r S_0 \frac{\partial w}{\partial S_0}  \ . 
\label{BS0}\end{equation}

\begin{figure}
\centerline{\includegraphics[height=4in]{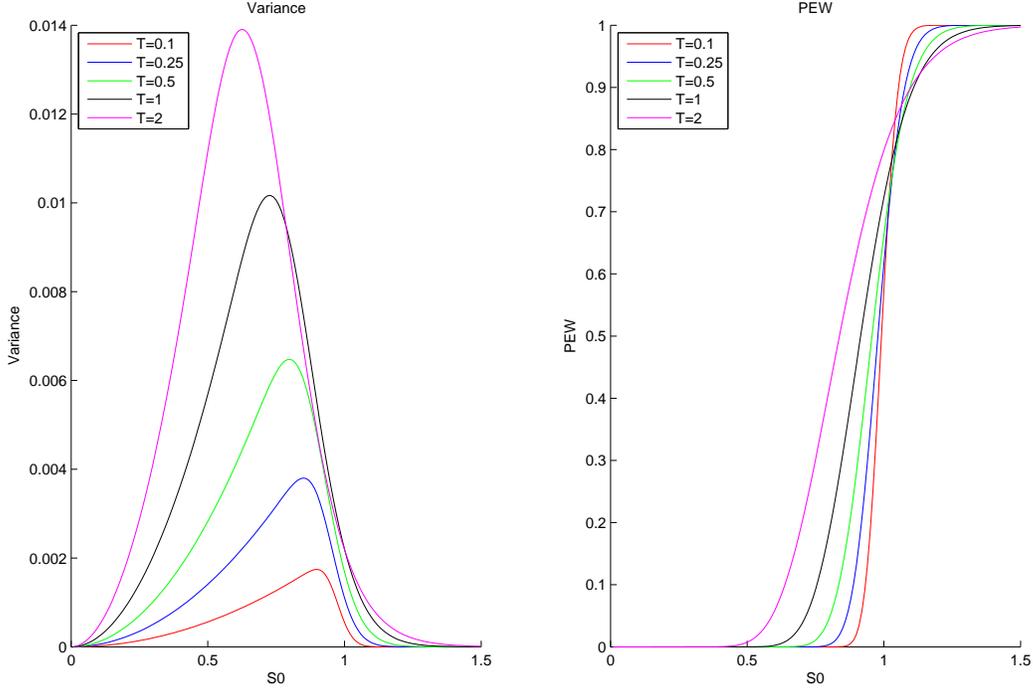}} 
\caption{Variance and PEW for a European put option with $r=0.1$, $\sigma=0.15$, $K=1$} 
\label{f1}\end{figure}

In Figure \ref{f1} we show the variance and PEW of a put option with $r=0.1$, $\sigma=0.15$, $K=1$ 
as a function of $S_0$ and for a variety of different times $T$. As expected, the variance is small for both 
low and high values of $S_0$, and in general increases with $T$. The location of the maximum variance falls
as $T$ increases. The PEW tends to a step function for small times. 

\section{A down-and-out put option}    

The variance and the PEW can also be computed analytcially for European barrier options. We 
illustrate this in this section by considering the case of a down-and-out put; that is, a 
European put option with the usual 5 parameters $r,\sigma,T,K,S_0$ and an additional ``barrier''
$B<{\rm min}(K,S_0)$; if at any time during the life of the option the price of the underlying 
falls below $B$, then the value of the option is knocked to zero (irrespective of how the price 
develops after hitting the barrier). 

We use the following result proven in \cite{etheridge2002course}: if $Y(t)=W(t)+bt$ is a Weiner 
process with drift then 
\begin{equation}
{\bf P}( {\rm min}_{0\le t\le T} Y(t) \le a, x\le Y(T)\le x+dx ) 
  =  \frac{A}{\sqrt{2\pi T}}\exp\left( -\frac{(x-bT)^2}{2T} \right)\ dx 
\label{eth}\end{equation} 
where 
$$ A = \left\{
\begin{array}{ll}
1 & a\ge x  ~{\rm or}~ a\ge 0 \\
\exp\left( \frac{2a(x-a)}{T}\right) & {\rm otherwise}
\end{array}
\right. \ .$$
Note that if $A=1$ the RHS in (\ref{eth}) is simply ${\bf P}( x\le Y(T)\le x+dx )$. If 
$a>0$ or $a\ge x$ then the requirement that ${\rm min}_{0\le t\le T} Y(t) \le a$ is automatically
satisfied, as $Y(t)$ starts from $0$ and ends at $x$. This is why $A=1$ in these cases. 
Remarkably the correction factor $A$ does not depend on $b$. 
It is straightforward to verify that the probability
density in (\ref{eth}), in both regions, 
satisfies the forward Kolmogorov equation 
$$ \frac{\partial f}{\partial T} = \frac12 \frac{\partial^2 f}{\partial x^2} 
  - b \frac{\partial f}{\partial x}\ . $$ 

The current value of a put option can be written in the form
$$ P = e^{-rT}\left( K - S_0e^{\sigma Y(T)} \right)_+ $$ 
where $Y(T)$ is a Weiner process with drift $b=\frac{r}{\sigma}-\frac{\sigma}{2}$. 
To avoid hitting the barrier we need ${\rm min}_{0\le t\le T}Y(t) \ge a$ where 
$a=\frac1\sigma\ln\left(\frac{B}{S_0}\right) < 0 $. To obtain a nonzero return 
from the option we need $Y(T)\le\frac1\sigma\ln\left(\frac{K}{S_0}\right)$.  
Putting this all together we deduce that in the presence of the knock-out barrier at $B$
\begin{eqnarray*}
PEW(P) &=&   1- \int_a^{\frac1\sigma\ln\left(\frac{K}{S_0}\right)} 
  \frac{1-\exp\left( \frac{2a(x-a)}{T} \right) }{\sqrt{2\pi T}}
   \exp\left( - \frac{(x-bT)^2}{2T} \right) \ dx
\\
{\bf E}[P] &=&  
\int_a^{\frac1\sigma\ln\left(\frac{K}{S_0}\right)}  e^{-rT} \left( K - S_0e^{\sigma x} \right)
  \frac{1-\exp\left( \frac{2a(x-a)}{T} \right) }{\sqrt{2\pi T}}
   \exp\left( - \frac{(x-bT)^2}{2T} \right) \ dx
 \\
{\bf E}[P^2] &=& 
\int_a^{\frac1\sigma\ln\left(\frac{K}{S_0}\right)}  e^{-2rT} \left( K - S_0e^{\sigma x} \right)^2 
  \frac{1-\exp\left( \frac{2a(x-a)}{T} \right) }{\sqrt{2\pi T}}
   \exp\left( - \frac{(x-bT)^2}{2T} \right) \ dx
\end{eqnarray*}
where we have written 
$$ a = \frac1\sigma\ln\left(\frac{B}{S_0}\right) ~~{\rm and}~~ 
b=\frac{r}{\sigma}-\frac{\sigma}{2} \ . $$ 
Throughout we are assuming $B<{\rm min}(K,S_0)$.  
All the integrals are Gaussian, and the calculations are straightforward though tedious. 
The final results are as follows: 
\begin{equation}
PEW(P) = 1 - \left( \Phi(Q_0)-\Phi(Q_1) +  \left( \frac{B}{S_0} \right)^{\frac{2r}{\sigma^2}-1} 
  \left( \Phi(Q_2)-\Phi(Q_3)  \right)   \right) \ ,  
\label{DAO0}\end{equation}
where 
\begin{eqnarray*}
Q_0 &=& \frac1{\sigma\sqrt{T}} \left(  \ln\left( \frac{S_0}{B} \right) + \left( r-\frac12\sigma^2 \right)T \right) \ ,\\
Q_1 &=& \frac1{\sigma\sqrt{T}} \left(  \ln\left( \frac{S_0}{K} \right) + \left( r-\frac12\sigma^2 \right)T \right) \ ,\\
Q_2 &=& \frac1{\sigma\sqrt{T}} \left(  \ln\left( \frac{B^2}{KS_0} \right) + \left( r-\frac12\sigma^2 \right)T \right) \ ,\\
Q_3 &=& \frac1{\sigma\sqrt{T}} \left(  \ln\left( \frac{B}{S_0} \right) + \left( r-\frac12\sigma^2 \right)T \right) \ . 
\end{eqnarray*}
\begin{eqnarray}
{\bf E}[P] &=& Ke^{-rT} \left( \Phi(Q_0)-\Phi(Q_1) +  \left( \frac{B}{S_0} \right)^{\frac{2r}{\sigma^2}-1} 
  \left( \Phi(Q_2)-\Phi(Q_3)  \right)   \right)  \nonumber\\ 
&& -S_0 \left( \Phi(Q_4)-\Phi(Q_5) +  \left( \frac{B}{S_0} \right)^{\frac{2r}{\sigma^2}+1} 
  \left( \Phi(Q_6)-\Phi(Q_7)  \right)   \right) \label{DAO1} 
\end{eqnarray}
where 
\begin{eqnarray*}
Q_4 &=& \frac1{\sigma\sqrt{T}} \left(  \ln\left( \frac{S_0}{B} \right) + \left( r+\frac12\sigma^2 \right)T \right) \ ,\\
Q_5 &=& \frac1{\sigma\sqrt{T}} \left(  \ln\left( \frac{S_0}{K} \right) + \left( r+\frac12\sigma^2 \right)T \right) \ ,\\
Q_6 &=& \frac1{\sigma\sqrt{T}} \left(  \ln\left( \frac{B^2}{KS_0} \right) + \left( r+\frac12\sigma^2 \right)T \right) \ ,\\
Q_7 &=& \frac1{\sigma\sqrt{T}} \left(  \ln\left( \frac{B}{S_0} \right) + \left( r+\frac12\sigma^2 \right)T \right) \ . 
\end{eqnarray*}
\begin{eqnarray}
{\bf E}[P^2] &=& 
K^2e^{-2rT}\left( \Phi(Q_0)-\Phi(Q_1) +  \left( \frac{B}{S_0} \right)^{\frac{2r}{\sigma^2}-1} 
  \left( \Phi(Q_2)-\Phi(Q_3)  \right)   \right)  \nonumber\\ 
&& -2KS_0e^{-rT} \left( \Phi(Q_4)-\Phi(Q_5) +  \left( \frac{B}{S_0} \right)^{\frac{2r}{\sigma^2}+1} 
  \left( \Phi(Q_6)-\Phi(Q_7)  \right)   \right) \nonumber\\
&& +S_0^2e^{\sigma^2 T}\left( \Phi(Q_8)-\Phi(Q_9) +  \left( \frac{B}{S_0} \right)^{\frac{2r}{\sigma^2}+3} 
  \left( \Phi(Q_{10})-\Phi(Q_{11})  \right)   \right) \label{DAO2} 
\end{eqnarray}
where 
\begin{eqnarray*}
Q_8 &=& \frac1{\sigma\sqrt{T}} \left(  \ln\left( \frac{S_0}{B} \right) + \left( r+\frac32\sigma^2 \right)T \right) \ ,\\
Q_9 &=& \frac1{\sigma\sqrt{T}} \left(  \ln\left( \frac{S_0}{K} \right) + \left( r+\frac32\sigma^2 \right)T \right) \ ,\\
Q_{10}&=& \frac1{\sigma\sqrt{T}} \left(  \ln\left( \frac{B^2}{KS_0} \right) + \left( r+\frac32\sigma^2 \right)T \right) \ ,\\
Q_{11} &=& \frac1{\sigma\sqrt{T}} \left(  \ln\left( \frac{B}{S_0} \right) + \left( r+\frac32\sigma^2 \right)T \right) \ . 
\end{eqnarray*}
Formulae (\ref{DAO0}), (\ref{DAO1}) and (\ref{DAO2}) 
are of limited importance in themselves. Their significance lies 
more in the fact that, as can be readilly verified using a symbolic manipulator, {\em Formulae (\ref{DAO0}), 
(\ref{DAO1}) and (\ref{DAO2}) satisfy the (modified) Black-Scholes equations (\ref{BS0}), (\ref{BS1}) and (\ref{BS2}) 
respectively}. As we shall see in the next section, this can provide a 
numerical procedure to determine the quantities we need even when
analytic expressions are not available.

\begin{figure}
\centerline{\includegraphics[height=4in]{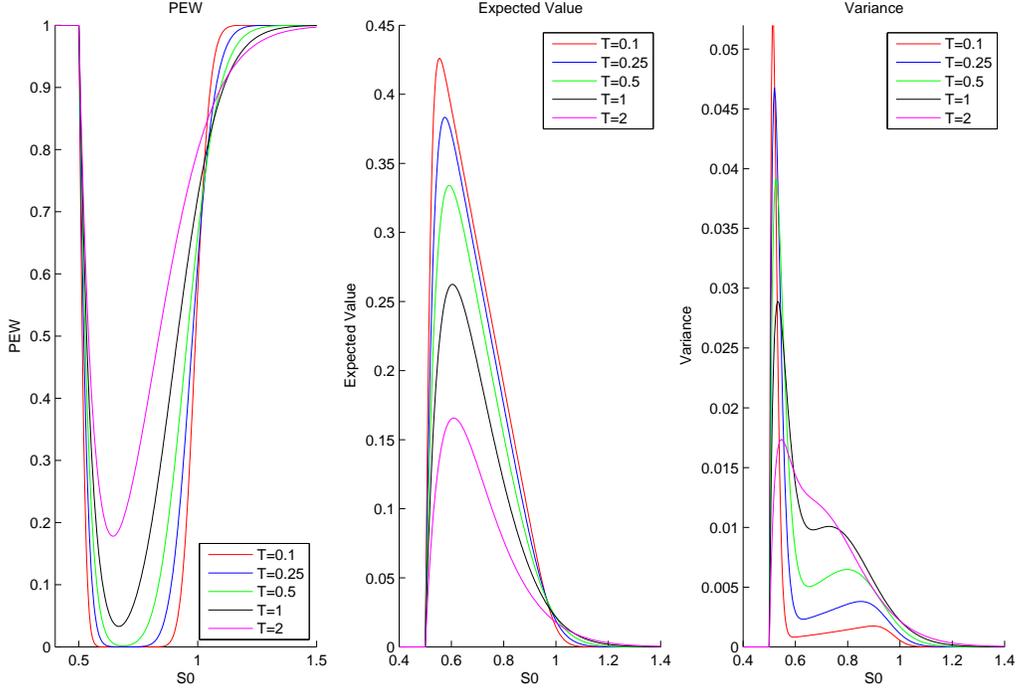}} 
\caption{PEW, expected value and variance for a down-and-out barrier 
put option with $r=0.1$, $\sigma=0.15$, $K=1$, $B=0.5$.} 
\label{f3}\end{figure}

We conclude this section by presenting plots of the PEW, expected value and variance  of a down-and-out put option
with the same parameters as the European option considered in section 2, but with a barrier at $B=0.5$. See Figure
\ref{f3}. For $S_0$ well above the barrier the results are similar to those for the option without a barrier. 
But as $S_0$ decreases towards the barrier, the PEW increases towards $1$, the expected value drops and the variance
increases rapidly, before dropping again to $0$ when $S_0=B$. 

\section{American options}  

For the case of American options there is no analytic expression for the Black-Scholes price,
and we do not expect there to be analytic expressions for the variance or the PEW. (We 
limit the discussion here to American puts on non-dividend paying stocks; if no dividends
are paid then for call options there is no added value in the early exercise feature 
\cite{wilmott1995mathematics}. The 
discussion can be extended to the case of continuous-dividend paying stocks.) The preferred 
approach for pricing American options (on a single underlying asset) is by numerical solution 
of a free boundary problem for the Black-Scholes PDE. For an option with strike $K$ and 
expiration $T$ on an underlying with volatility $\sigma$, the Black-Scholes price $u(s,t)$ 
(as a function of the underlying price $s$ and time $t$) should satisfy the PDE 
\begin{equation}
\frac{\partial u}{\partial t} + 
\frac12\sigma^2 s^2 \frac{\partial^2 u}{\partial s^2}
+ r s \frac{\partial u}{\partial s} - r u  =0 
\label{BS1a}\end{equation}
on the domain  $0<t<T$, $s>b(t)$, where $b(t)$ is the unknown {\em early exercise price}. 
$u(s,t)$ should satisfy the Dirichlet boundary conditions 
\begin{eqnarray*}
\lim_{s\rightarrow\infty} u(s,t) &=& 0 \ , \\ 
u(b(t),t) &=& K - b(t) 
\end{eqnarray*}
and the final time condition 
$$
u(s,T) = 0 \ , \qquad s>b(T)  \ .
$$
The early exercise price $b(t)$ is determined by the {\em smooth pasting condition} 
$$
u_s(b(t),t) = -1  
$$
along with the final time condition $b(T)=K$. See \cite{wilmott1995mathematics}
for a full exposition. 

Completely analogously to our findings in the previous 2 sections, 
the expectation $v(s,t)$ of the square of the current value of the 
option should obey the modified Black-Scholes equation 
\begin{equation}
\frac{\partial v}{\partial t} + 
\frac12\sigma^2 s^2 \frac{\partial^2 v}{\partial s^2}
+ r s \frac{\partial v}{\partial s} - 2 r v  =0 
\label{BS2a}\end{equation}
on the same domain, 
with boundary conditions 
\begin{eqnarray*}
\lim_{s\rightarrow\infty} v(s,t) &=& 0 \ , \\ 
v(b(t),t) &=& (K - b(t))^2  
\end{eqnarray*}
and  final time condition 
$$
v(s,T) = 0 \ , \qquad s> b(T)  \ .
$$
These conditions enforce vanishing variance at all boundaries. Note that  
there is no reason to expect the  variance to have zero derivative 
at the early exercise boundary --- thus it need not paste smoothly onto 
the variance below the early exercise boundary, which vanishes. 

Similarly, the PEW $w(s,t)$ should obey 
\begin{equation}
\frac{\partial w}{\partial t} + 
\frac12\sigma^2 s^2 \frac{\partial^2 w}{\partial s^2}
+ r s \frac{\partial w}{\partial s}  =0 
\label{BS0a}\end{equation}
with boundary conditions 
\begin{eqnarray*}
\lim_{s\rightarrow\infty} w(s,t) &=& 1 \ , \\ 
w(b(t),t) &=& 0  
\end{eqnarray*}
and  final time condition 
$$
w(s ,T) = 1 \ .  \qquad s>b(T)  \ . 
$$
Note that for the PEW there is a discontinuity in the boundary data 
at $t=T$, $s=B(T)$. 

\begin{figure}
\centerline{\includegraphics[height=2.5in]{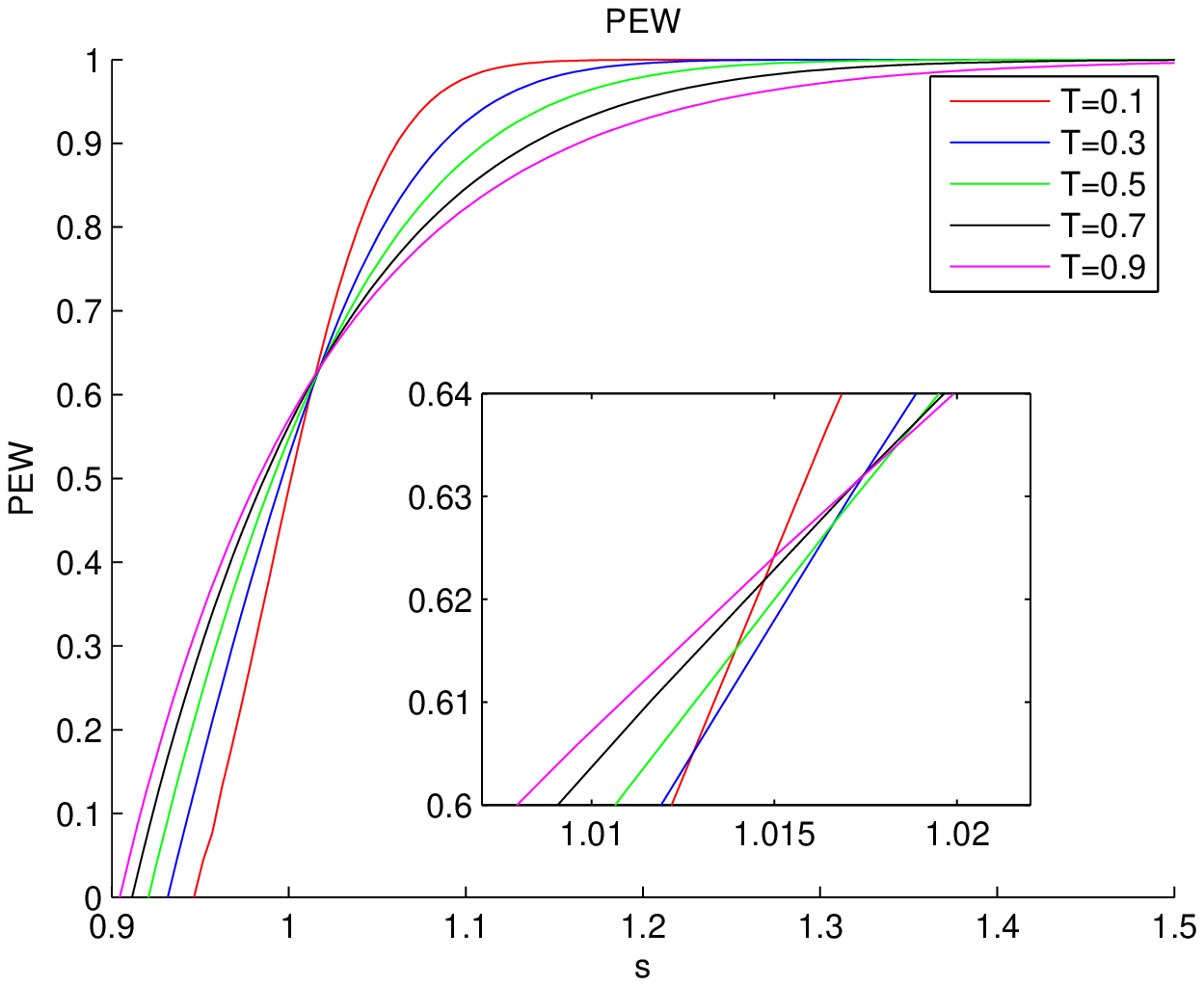}
            \includegraphics[height=2.5in]{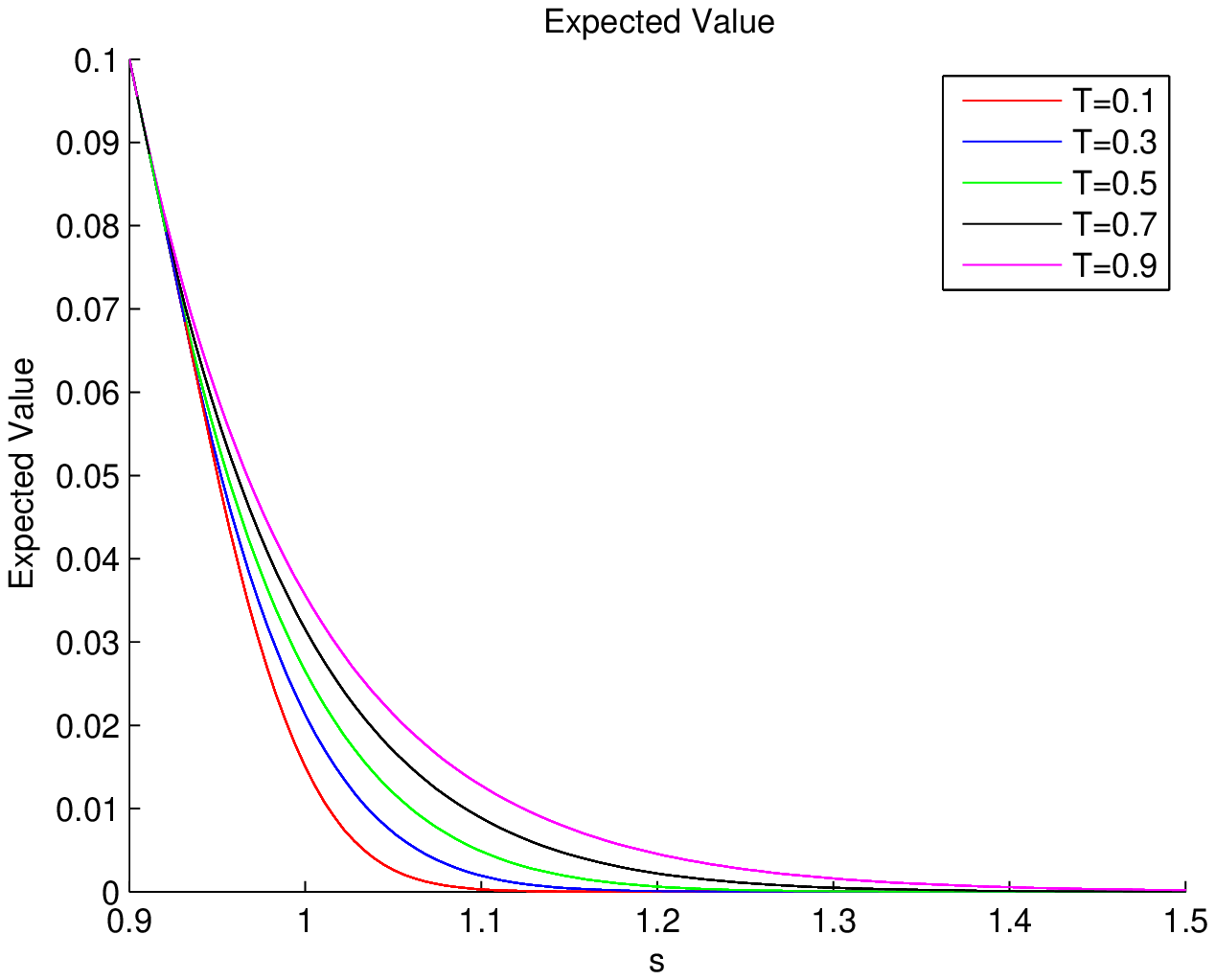}} 
\centerline{\includegraphics[height=2.5in]{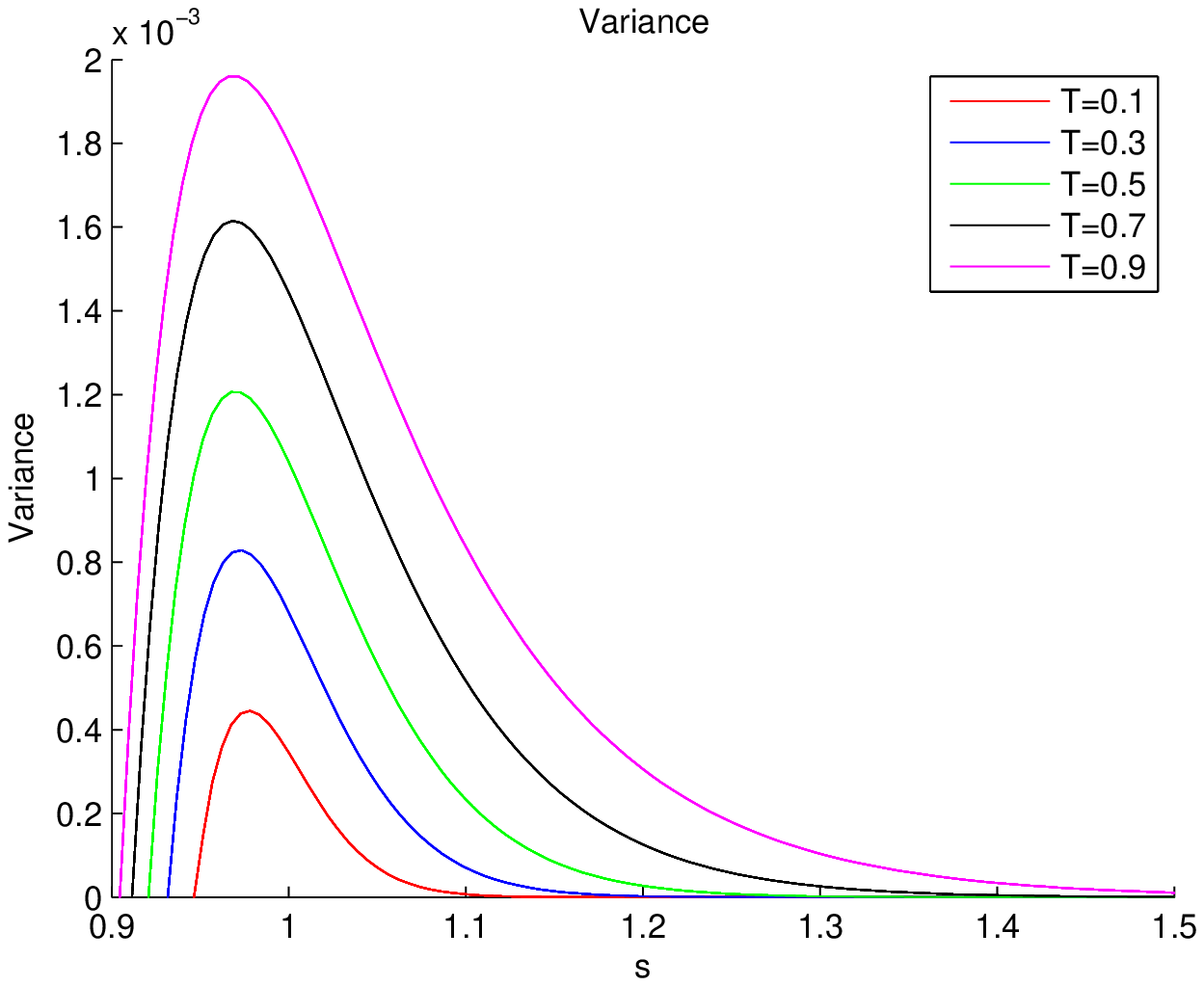}} 
\caption{PEW, expected value and variance for an American 
put option with $r=0.1$, $\sigma=0.15$, $K=1$.
Including a blow-up of the intersection region for the PEW. 
The PEW and variance go to zero on the early exercise boundary. 
} 
\label{f4}\end{figure}

Note that the early exercise price $b(t)$ is determined by the free boundary problem for the 
standard Black-Scholes equation (\ref{BS1a}), and this is then taken as input to solve the relevant 
problems for  (\ref{BS2a}) and  (\ref{BS0a}). In practice we solve all 3 PDEs using a front-fixing 
technique \cite{wu1997front,duffy2011finite}. Define new dimensionless coordinates $y,\tau$ via
$$  y = \log\left( \frac{s}{b(t)} \right) \ , \qquad \tau = \frac{\sigma^2}{2}(T-t)  $$
and rescale the unknown functions $u,v,w,b$ via
$$  u(s,t) = KU(y,\tau)  \ , \qquad
    v(s,t) = K^2V(y,\tau)  \ , \qquad
    w(s,t) = W(y,\tau)  \ , \qquad
    b(t) = K B(\tau)\ . $$
The system of equations becomes
\begin{eqnarray}
\frac{\partial U}{\partial \tau} &=&  \frac{\partial^2 U}{\partial y^2}
+ \left(  \frac{2r}{\sigma^2} - 1 + \frac{B'(\tau)}{B(\tau)}  \right)  \frac{\partial U}{\partial y} 
- \frac{2r}{\sigma^2} U \ ,  \label{BS1b} \\ 
\frac{\partial V}{\partial \tau} &=&  \frac{\partial^2 V}{\partial y^2}
+ \left(  \frac{2r}{\sigma^2} - 1 + \frac{B'(\tau)}{B(\tau)}  \right)  \frac{\partial V}{\partial y} 
- \frac{4r}{\sigma^2} V \ ,  \label{BS2b} \\ 
\frac{\partial W}{\partial \tau} &=&  \frac{\partial^2 W}{\partial y^2}
+ \left(  \frac{2r}{\sigma^2} - 1 + \frac{B'(\tau)}{B(\tau)}  \right)  \frac{\partial W}{\partial y} \ , 
\label{BS0b}
\end{eqnarray}
all on the fixed domain $0<\tau<\tau_{\rm max} = \frac12 \sigma^2 T $, $y>0$, with initial conditions 
$$ U(y,0) = V(y,0)=0\ ,  \qquad W(y,0) = B(0)=1 \ ,  $$ 
Dirichlet boundary conditons 
$$
U(0,\tau)=1-B(\tau)\ ,\qquad
V(0,\tau)=(1-B(\tau))^2\ ,\qquad
W(0,\tau)=0 
$$
and
$$
\lim_{y\rightarrow\infty} U(y,\tau) = \lim_{y\rightarrow\infty} V(y,\tau) = 0 \ , \qquad 
\lim_{y\rightarrow\infty} W(y,\tau) = 1 \ ,
$$
and finally the  Neumann condition 
$$
U_y(0,\tau)=-B(\tau)\ .
$$
We solve this system using the Crank-Nicolson scheme to advance $U,V,W$; the simplest first order discretization
is used to approximate $B'(\tau)$, and a 3-point, second order approximation is used for $U_y(0,\tau)$. 
At each time step the Newton-Raphson method is used to update $B$ in a manner that the Neumann boundary condition 
holds. Results are displayed in Figure \ref{f4} for an option $r=0.1$, $\sigma=0.15$, $K=1$. By construction the 
variance and PEW go to zero on the early exercise boundary. The variance apparently increases uniformly with $T$ 
(for fixed $s$), which makes good sense  --- but note this was not exactly the case for the European option 
(see Figure \ref{f1}). As $T$ tends to zero the PEW tends to a step function, as expected. 

A few notes on numerical aspects of the above calculations: The discontinuity in the
derivative of the PEW at the early exercise boundary did not interfere with 
convergence of the Crank-Nicolson method, but the usual $O(h^2)$ error behavior was 
reduced to $O(h)$. The variance calculation was verified using a Monte-Carlo technique;
once the early exercise curve has been calculated, standard Monte-Carlo techniques can 
be used to get the distribution (and in particular the mean and variance) of the return
from the option. These results confirmed the PDE calculations, but a ``barrier correction'' 
following \cite{broadie1997continuity}
was needed to compensate for discrete-time observation of whether the stock hits the 
early exercise price in the Monte-Carlo simulation.  There are also Monte-Carlo techniques 
for pricing American options that do not depend on knowing the early exercise price, 
for example the method of \cite{broadie1997pricing,broadie1997enhanced} uses a tree approach to 
compute two estimates which are lower and upper bounds for the option price. The variances
of the lower and upper bounds computed in this approach were found to be substantially lower than the 
option variance computed by PDE methods. This is perfectly reasonable, the choices made to obtain 
bounds eliminate a lot of the variation of the option price. 

\section{Variance and PEW in Investment Decisions.}   

Consider an investor buying a put option on a stock. 
The investor faces the problem of selecting from a universe of 
options indexed by their strike $K$ and time to expiration $T$, with prices increasing as 
functions of both $K$ and $T$. Assuming the options are priced by the classical Black-Scholes 
formula (for European puts), how is the investor to make this decision? 

\begin{figure}
\centerline{\includegraphics[height=4in]{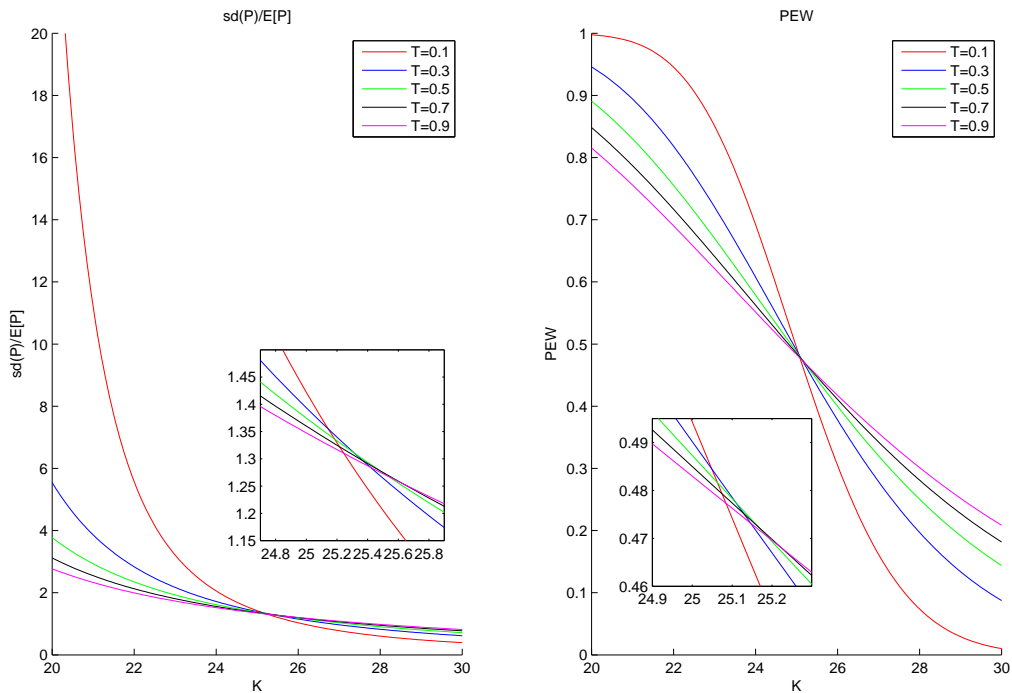}} 
\caption{${\rm sd}(P)/{\bf E}[P]$ and PEW for European put options on a stock 
with $r=0.02$, $\sigma=0.25$, $S_0=25$. Includes blow-ups of the intersection regions.}
\label{f2}\end{figure}

The expected rate of return does not depend which options the investor buys
(and for that matter is the same as that for a direct investment in the stock). Ignoring 
possible external constraints on time frames and margin requirements, this question is entirely a 
question of the risk profile the investor wishes to assume. As a first quantitative measure of 
the risk we would suggest to look at ${\rm sd}(P)/{\bf E}[P]$, i.e. the standard
deviation of the option value per unit price. In the first graph in 
Figure \ref{f2} we plot this, for a stock 
with $r=0.02$, $\sigma=0.25$, $S_0=25$, as a function of $K$ and for a range of values of 
$T$. As we might expect, out-of-the-money options ($K<S_0$) are apparently  much more risky than 
in-the-money options ($K>S_0$), and whereas for deep in-the-money options shorter term 
means lower risk, for deep out-of-the-money shorter term means greater risk. (The behavior 
close-to-the-money is interesting, as the insert in the figure  illustrates.) 

From the exact formulae for the variance of the European put option it can be shown 
that as $K$ tends to infinity the variance ${\rm Var}(P)$ tends to 
$S_0^2(e^{\sigma^2 T}-1)$, which is the variance of the stock price 
(discounted to current values). Thus the first plot in Figure \ref{f2} shows the 
evidently sensible result that for fixed
$T$ the risk associated with buying put options decreases with the strike $K$, ultimately
decreasing to the risk of the stock itself. 

Another quantity the investor may wish to look at is the PEW. 
This is also displayed in Figure \ref{f2}. In fact we see that PEW behaves very similarly to 
${\rm sd}(P)/{\bf E}[P]$.  Indeed, by Chebyshev's inequality 
$$ {\bf P}( |P - {\bf E}[P]| \ge {\bf E}[P] ) ~\le~ \frac{{\rm Var}(P)}{\left( {\bf E}[P] \right)^2 } $$ 
and thus 
$$ PEW ~=~ {\bf P}(P=0) ~\le~  
{\bf P}( |P - {\bf E}[P]| \ge {\bf E}[P] ) ~\le~ 
\left( \frac{{\rm sd}(P)}{ {\bf E}[P] }\right)^2 \ . $$
In practice, as usual with Chebyshev's inequality, this is not a very useful inequality
quantitatively, but it does, however, correctly indicate that the general trends of PEW and 
${\rm sd}(P)/{\bf E}[P]$ are similar. 

It should be emphasized that the risk we are discussing here is risk as viewed from the buyer's side.
For PEW this is clear; for the buyer of the option, high PEW means a high probability of losing 
100\% of the investment, which is an uncontroversial, if rather naive, measure of risk. 
For the seller the opposite is true; high PEW is a measure of safety for the seller. This reflects 
the general asymmetry between buyer and seller in matters of risk --- a buyer prefers to buy low
risk, a seller prefers to sell away high risk. For ${\rm sd}(P)/{\bf E}[P]$ things are a little less
clear. If we are discussing an options trader, who is buying (or selling) an option, say, to add to 
(or remove from) a portfolio, then indeed, ${\rm sd}(P)/{\bf E}[P]$, which 
measures the spread of the return from the option, is a measure of buyer's risk, and the 
same asymmetry between buyer and seller exists. (Note, however, that if the trader holds other 
options on the same underlying then it is of course a mistake 
to consider just the risk on an individual asset in the portfolio and ignore correlations.)
The potentially confusing situation comes in considering an investor who wishes to 
write, say, a single in-the-money put option, with a view to actually owning the stock
at a later date when the option is assigned. Despite being a seller of the option, this
investor may well prefer {\em lower} ${\rm sd}(P)/{\bf E}[P]$. For this seller, the sale 
does not eliminate risk, but adds to it --- so the considerations of a buyer may be more 
appropriate. 

\section{Risk-averse/Risk-seeking Option Pricing.}   

\begin{figure}
\centerline{\includegraphics[height=4in]{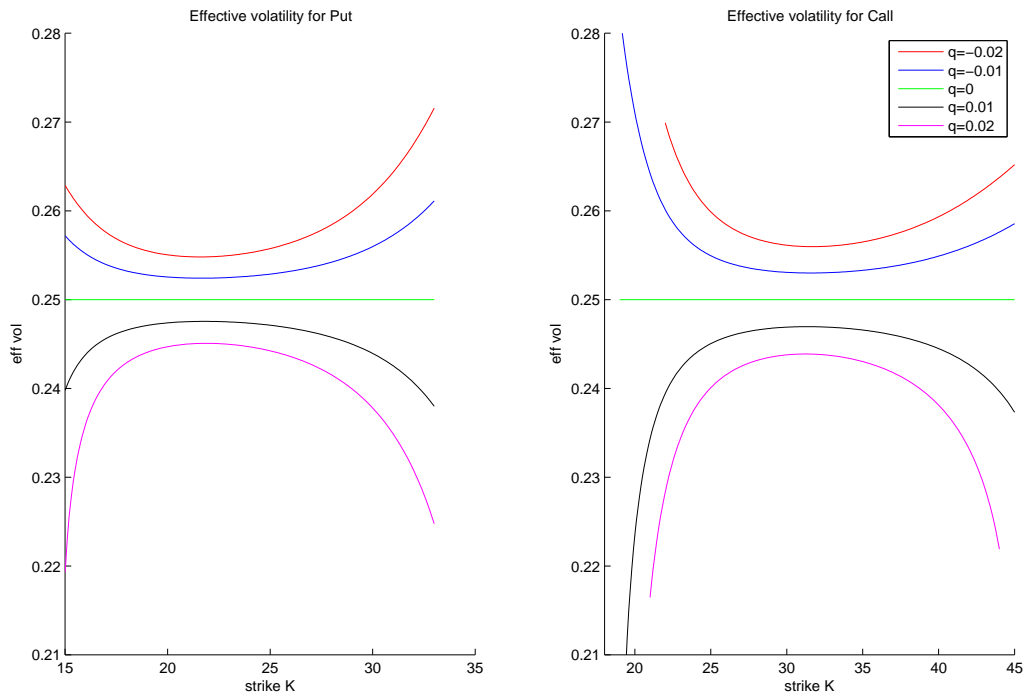}} 
\caption{Effective volatility of European call and put options priced using (\ref{pf}),  
as a function of $K$. Values of $q$ from top to bottom: $-0.02,-0.01,0,0.01,0.02$.  
Stock parameters $r=0.02$, $\sigma=0.25$, $S_0=25$; option expiration $T=0.5$.}
\label{f5}\end{figure}

The previous section concerns the use of risk measures of options in individual investment 
decisions. In this section we consider the following question: Is it possible that the 
market incorporates a measure of risk in option prices? The simplest imaginable way to
incorporate risk premium is to consider a pricing formula of the form 
\begin{equation}
{\rm option~price} = {\rm expected~value} - q ~ \times ~ {\rm standard~deviation} 
\label{pf}\end{equation}
where $q$ is a parameter quantifying the average risk aversion of option buyers. 
Since in the classical Black-Scholes formulas option prices are monotonic increasing functions of 
volatility, we can translate the above change in option prices into 
effective volatility values.  Note that for $q>0$ (net risk aversion) the above formula gives
prices below classical Black-Scholes, and hence effective volatility is lower than input volatility. 
For $q<0$ (net risk seeking) the reverse is true. In Figure \ref{f5} we plot the 
effective volatility of 
European call and put options as a function of $K$ for a few 
values of $q$. We assume an underlying with $r=0.02$, $\sigma=0.25$, $S_0=25$ and look
at options with expiration $T=0.5$. 

We see that for $q<0$ our model predicts a volatility smile and for $q>0$ a volatility frown. 
The plots in Figure \ref{f5} are somewhat misleading, as very large ranges of strikes 
are shown (differing in the 2 subplots). For deep out-of-the-money and deep-in-the-money 
options, the classical Black-Scholes price is very insenstive to the volatility. Thus a small
change in price requires a substantial change in volatility. This is the reason for the sharp 
upturns and downturns in effective volatility, particularly noticeable in the case of the 
call. Nevertheless, it seems that Equation (\ref{pf}) provides a 1-parameter extension of 
the Black-Scholes model which can explain a volatility smile. The standard extensions ---
local volatility models 
\cite{derman1994riding,dupire1994pricing} 
and stochastic volatility models
\cite{hull1987pricing,heston1993closed,hagan2005probability}
 --- typically need at least 
3 parameters. 

The term structure of effective volatility implied by Equation (\ref{pf}) is complicated
and we do not describe it here. 
However, we note that since $q$ is a parameter quantifying risk aversion, it is quite 
reasonable to take $q$ to depend on the timeframe $T$, thereby allowing any desired 
term structure to be built into the model. 
As of yet we have not tested to see whether the model can provide a reasonable fit 
to data. 

\section{Concluding remarks} 

In this paper we have explained methods to calculate the variance (and the PEW) of standard options, 
and outlined potential applications. Despite the fact that we have not yet performed comprehensive
empirical studies, we think the potential usefulness of the variance and/or the PEW is self-evident. 
A further point may help to make this clear: In the classical Black-Scholes option formulas, the 
price of an option depends on 5 parameters $S_0,K,T,r,\sigma$. By rescaling units of time 
and money 2 of these parameters can be eliminated. By moving to a frame in which all money accumulates
at a rate $r$, dependence on $r$ can be eliminated. Thus there is only non-trivial dependence on 
2 effective parameters, which are typically taken to be the combinations 
$\sigma\sqrt{T}$ and $\frac{S_0}{Ke^{-rT}}$. If an option is priced using the classical Black-Scholes
formulas, then two pieces of data are being compressed into one --- in other words, information 
is being lost. By also looking at the variance we take all information into account. 

In addition to empirical studies and investigation of potential applications, 
there are many extensions of the current work that come to mind,
and we mention just a few:
\begin{itemize}
\item The variance formulas for European options 
in this paper should be extended to cover the case of options on 
(continuous) dividend-paying stocks, for which exact formulas should still be available. 
\item  For American options, we have seen exact formulas are not available, but we believe
it should be easy to provide good approximations, particularly in the limit of low 
volatility, following the work of Widdicks et al. \cite{widdicks2005black,duck2009singular}
\item Similarly for options on multiple assets, asymptotic approximations should be 
available. Good methods for options on multiple assets should involve incoporating 
asymptotic results into numerical schemes to have the best of both worlds. 
\end{itemize}

\vskip.2in 

\noindent{\bf Acknowledgements.} 
We thank Yaniv Zaks and Evgenia Apartsin for useful conversations. 
This work includes material included in the M.Sc. thesis of the first author. 

\vskip.2in 

\bibliographystyle{acm} \bibliography{all}
\end{document}